\documentclass{article}
\usepackage{amsmath,amsxtra,amssymb,latexsym, amscd,amsthm}
\usepackage{indentfirst}
\usepackage[mathscr]{eucal}
\usepackage[pctex32]{graphics}
\usepackage {graphicx,epsf}
\def\bfr{{\bf r}}
\def\bfv{{\bf v}}

\def\rr{\varrho}

\def\bfx{{\bf x}}

\def\bfz{{\bf z}}
\def\sqb{\sqsubseteq}
\def\sqp{\sqsupseteq}

\def\pa{\partial}

\def\rr{\rho}

\newcommand{\beq}{\begin{equation}}
\newcommand{\eeq}{\end{equation}}
\newcommand{\beqn}{\begin{eqnarray}}
\newcommand{\eeqn}{\end{eqnarray}}
\newcommand{\beqnn}{\begin{eqnarray*}}
\newcommand{\eeqnn}{\end{eqnarray*}}
\newcounter{A}
\begin{document}
\title{Canonical description of incompressible fluid -- Dirac brackets
approach}
\author{Sonnet H. Q. Nguyen \footnote{e-mail: sonnet@cft.edu.pl, corresponding author.} , \L ukasz A. Turski
\footnote{e-mail: laturski@cft.edu.pl}  \\
Center for Theoretical Physics, Polish Academy of Science, and \\
College of Science. Al. Lotnik\'ow 32/46. 02-668 Warszawa. Poland}
\date{May 05, 1999} 
\markboth{}{{Sonnet H. Q. Nguyen, \L. A. Turski, Canonical description of incompressible fluid ...}}
\maketitle
\pagestyle{myheadings}
\noindent{\bf Abstract.-} We present a novel canonical description of
the incompressible fluid dynamics. This description uses the dynamical
constraints, in our case reflecting incompressibility assumption, and
leads to replacement of usual hydrodynamical Poisson brackets for
density and velocity fields with Dirac brackets. The resulting
equations are then known nonlinear, and nonlocal in space, equations
for  incompressible fluid velocity.\\

\vskip0.4cm 
\leftskip0.05cm 
{\it Keywords}: Fluid dynamics, Poisson structure, Dirac brackets, Canonical description, 
                         Hamiltonian formulation, Dirac constraints. \\
This is a draft of an article published in:  Physica {\bf A 272}, 48-55, (1999).
\vskip0.4cm 
\leftskip0.05cm 
\section{Introduction}
\label{sec:introduction}
Canonical description of classical, compressible, isothermal fluid has
been developed in the past \cite{canbiblio,Langer,Kim} for various
purposes, for example description of superfluid${\ }^{4}$He
\cite{canbiblio} or in kinetics of the first order phase
transformations \cite{Langer,Kim}. An attempt to extend this
formulation for the adiabatic flows has been proposed \cite{Barbara}
and used to analyze dynamical properties of thermally driven flows.
The isothermal flow canonical description can be generalize for the
case of viscous fluids \cite{Enz}  within the framework of the
metriplectic dynamics \cite{Turski1}. The Madelung representation for
the wave function results in hydrodynamic like picture of quantum
mechanics, where the ``only'' differences from Euler equations are
hidden in the quantum pressure term, which is proportional to
$\hbar^{2}$,  and in the quantization of the circulation, $\Gamma
=n(\hbar/m)$. Apart of that the canonical description of the quantum
fluid is then identical to that of the classical one. The dissipative
generalization of the Schr{\"o}dinger equation \cite{Gisin}  also
allows for metriplectic interpretation, which differs from the
classical one \cite{Turski2}.

The fundamental point in all of the above listed formulations of fluid
dynamics is that the fluid density $\rr$ is one out of the pair of
canonically conjugated variables. For potential flows the other
canonical variable is the velocity potential $\phi$. In case of
general flow the  two  additional Clebsh potentials $\lambda, \mu$
\cite{Clebsh} in the velocity field representation $\bfv=-\nabla
\phi-\lambda\nabla\mu$ are canonically conjugated to each other. None
of these  descriptions can be applied to the case of incompressible
fluids.

The canonical description of incompressible flow is of considerable
importance, for example in formulation of statistical mechanics of
turbulent flow \cite{Enz}. Many attempts to provide such a canonical
formalism \cite{Arnold} failed to do so. The other important point is
that real turbulent flow are hardly incompressible, and the issues
like compressibility corrections to scaling laws  for turbulent flows
are still open \cite{Lvov}.

In this paper we propose a novel formulation of a canonical
description of the incompressible fluid based on the concept of the
Dirac brackets \cite{Dirac1}. The mathematical introduction to this
formalism valid for general dynamic system subject to some set of
constraints $\{\Theta_{a}=0, a=1\ldots N\}$ can be found in
\cite{Marsden1}. Dirac bracket approach to description of
incompressible membranes, within Lagrangian coordinates formulation of
continuous mechanics of membranes, was given in \cite{David}. We are
unaware of any other application of that formalism to continuum
mechanics problems. In separate publication we shall present other
application of the Dirac constraints formalism in classical mechanics
\cite{Sonnet2}.

The Dirac brackets, for incompressible fluid, are presented, in what
follows, within the Poisson bracket formulation of fluid mechanics
\cite{Langer,Kim,Enz}, which avoids cumbersome introduction of the
Clebsh potentials. Thus the state of fluid is described by specifying
its density and velocity fields.
%
\section{Compressible fluid}
Consider infinite 3-dimensional volume of the isothermal fluid with
density $\varrho({\bf r}, t)$ and velocity $ {\bf v}({\bf r}, t)$.
The Hamiltonian for such a system is given as:
\begin{eqnarray}\label{Ham1}
  {\cal H}\{\varrho,{\bf v}\}=\int\left[ \varrho {\bf v}^2/2
  +f(\varrho)\right] d^{3}r\;,
\end{eqnarray}
where $f(\varrho$) is the fluid Helmholtz free energy per unit volume,
related to the fluid pressure by
\begin{eqnarray}\label{press1}
  p=\varrho\frac{\partial f}{\partial \varrho} - f(\varrho) .
\end{eqnarray}

The Poisson bracket relations between fields $\varrho({\bf r}, t)$ and
$ {\bf v}({\bf r}, t)$ are \cite{Enz}:
\begin{eqnarray}\label{Poisson1}
   \{\varrho({\bf x},t), \varrho({\bf y},t) \} &=& 0\;, \nonumber\\
     \left\{ \varrho({\bf x},t), v^i({\bf y},t) \right\} &=&
      - \frac{\partial}{\partial {\bf x}^i}
     \delta({\bf x}-{\bf y})\;,\nonumber \\
     \left\{ v^i({\bf x},t), v^j({\bf y},t) \right \} &=&
      \delta({\bf x}-{\bf y})
           \frac{1}{\varrho({\bf x},t)} \epsilon^{ijk}
           (\nabla\times{\bf v})_k ({\bf x},t)\;.
\end{eqnarray}
The continuity equation is obtained by evaluating the Poisson bracket
$\{\varrho,{\cal H}\}$, and the Euler equation by $\{\bfv, {\cal H\}}$
\begin{eqnarray}\label{Euler}
  \partial_t\varrho(\bfr,t)&=&\{\varrho, {\cal H}\}=
  -\nabla\cdot\varrho\bfv\;,\nonumber\\
  \partial_t\bfv(\bfr,t)&=&\{\bfv,{\cal H}\}=
  -\bfv\cdot\nabla\bfv-(1/\varrho)\nabla p(\varrho)\;.
\end{eqnarray}
The above formulation of fluid mechanics can be derived from the least
action principle provided we choose the proper lagrangian. As shown by
Thellung \cite{canbiblio} this lagrangian density is the local
pressure.

The incompressible fluid, although an obvious simplification, is
adequate for all the flows when the local Mach number is small. The
incompressibility condition then is that the density
$\varrho(\bfr,t)-\varrho_0=0$ what also implies that
$\nabla\cdot\bfv=0$. Within the canonical formulation framework both
these conditions are regarded as Dirac constraints\cite{Dirac1}
$\Theta_a(\bfr,\bfv,t)=0$, $a=1,2$. Next section contains a brief
overview of the Dirac brackets theory.
%
\section{Dirac Brackets}

The definition of the Dirac brackets we shall use in the following is
a natural generalization for the original construction proposed by
Dirac \cite{Dirac1} and discussed in detail in \cite{Marsden1}. When
the physical  system with  phase space  $\cal{P}$ is subject to a set
of constraints $\{\Theta_{a}=0\}$ then its motion proceeds on a
submanifold  ${\cal P} \supset{\cal S}= \bigcup_{a=1}^{N}\{z \epsilon
{\cal P}|\Theta_{a}(z)=0\}$. If the Poisson bracket for two arbitrary
(sufficiently smooth etc.,) phase space functions $F$ and $G$ was
$\{F,G\}$, then the Dirac bracket $\sqsubseteq F,G \sqsupseteq$ is
defined as:
\begin{eqnarray}\label{Dirbrac}
  \sqsubseteq F,G \sqsupseteq= \{F,G
  \}=-\sum_{a,b}^{N}\{F,\Theta_{a}\}M_{ab}\{\Theta_{b},G\}\;,
\end{eqnarray}
where $M_{ab}$ is the inverse of the constraints Poisson bracket
matrix $W_{ab}=\{\Theta_{a},\Theta_{b}\}$.

The generalization of the Dirac bracket to the case of continuous
variables, like in hydrodynamics, is straightforward. The sum over the
indices $a$ is replaced by sum and integration over the space
variables and the inverse of the matrix
$W_{ab}(\bfr,\bfr')=\{\Theta_{a}(\bfr),\Theta_{b}(\bfr')\}$ is defined
as:
\begin{eqnarray}\label{inverse}
  \sum_{c}\int d\bfr'  W_{ac}(\bfr,\bfr')
  M_{cb}(\bfr',\bfr'')=\delta_{ab}\delta(\bfr-\bfr'')\;.
\end{eqnarray}
Dirac brackets, given by Eq.(\ref{Dirbrac}) replace the original
Poisson brackets in the equation of motion for the constrained system.
Thus for a phase space function $F$ the time evolution on the
submanifold ${\cal S}$ is governed by:
\begin{eqnarray}\label{moitondir}
  \left({\frac{\partial F}{\partial t}}\right)_{\cal S}=
  \sqsubseteq F,{\cal H}\sqsupseteq\;,
\end{eqnarray}
where ${\cal H}$ is system Hamiltonian. Next section will contain
application of the Dirac brackets to the description of the
incompressible fluid.
%
\section{Dirac Brackets for Incompressible Fluid}
The constraints used in constructing the incompressible fluid dynamics
are:
\begin{eqnarray}\label{constraints}
  \Theta_{1}&\equiv&\varrho(\bfr)-\varrho_{0}=0\;,\nonumber\\
  \Theta_{2}&\equiv&\nabla\cdot\bfv(\bfr)=0\;.
\end{eqnarray}
The constraints Poisson bracket matrix $W_{ab}(\bfr,\bfr')$ can be
evaluated using Eq.(\ref{Poisson1}), and it reads:
\beqn \label{wdirflu}
   W_{ab}(\bfr,\bfr') =\nabla_{r}^i\nabla_{r}^j\left(
    \left[\begin{array}{cc}
                0                \:   &  \:
              -\delta^{ij}\\
               \delta^{ij}, \:  &  \:
                              \left[\frac{1}{\varrho(\bfr)}\varepsilon^{ijk}
               (\nabla\times\bfv(\bfr))^k \right]
          \end{array}
   \right]\delta(\bfr-\bfr')\right)\;.
\eeqn

In the Dirac formalism  one needs  the inverse of the matrix
$W_{ab}(\bfr,\bfr')$ defined in(\ref{inverse}). The matrix elements
$M_{ab}(\bfr,\bfr')$ obey the set of partial differential equations,
written explicitly in the Appendix~A.  Solving these equations we find
matrix $M_{ab}(\bfr,\bfr')$ in the form:

\begin{eqnarray}\label{Mab1}
  M_{ab}(\bfr,\bfr') =
    \left[\begin{array}{cc}
               {\mathcal{M}}\{G\}  ,         \:   &  \:
              -G(\bfr-\bfr')\\
              G(\bfr-\bfr'), \:  &  \:0

          \end{array}
   \right]\;,
\end{eqnarray}
 where $G(\bfr-\bfr')= \mid \bfr -\bfr'\mid/4\pi$ is the Green
function for the Laplace operator in the infinite volume, and
\begin{eqnarray}\label{M11}
  {\mathcal{M}}\{G\}= -\int d\bfx'G(\bfr-\bfx')\nabla_{\bfx'}\cdot\left[
  \frac{1}{\varrho(\bfx')}\nabla_{\bfx'}G(\bfx'-\bfr')
  \times(\mathbf{\nabla\times\bfv(\bfx')})\right]\;.
\end{eqnarray}

%
\section{Dirac equations of motion for incompressible fluid}

Using the definition and the explicit form of the Dirac brackets given
in previous section and in the Appendix A, we first calculate the
Dirac bracket $\sqsubseteq \varrho , {\cal H}\sqsupseteq$. From the
definition (\ref{Dirbrac}), and (\ref{moitondir}) we obtain:

\begin{eqnarray}\label{Dirdens}
  \frac{\partial \varrho(\bfr,t)}{\partial t}=\sqsubseteq
  \varrho(\bfr,t), {\cal H}\sqsupseteq=\{\varrho(\bfr,t), {\cal
  H}\}-\sum_{ab}\int d{\bf z} d{\bf z}'\{\varrho(\bfr,t),\Theta_{a}({\bf
  z})\}M_{ab}({\bf z},{\bf z}')\{\Theta_{b}({\bf z}'),{\cal H}\}\;.
\end{eqnarray}
Explicit evaluation of the right hand side of Eq.(\ref{Dirdens}) is a
bit tedious, but using results from the Appendix~A and (\ref{Mab}) one
finds that it vanishes. Thus the continuity equation for
incompressible fluid, within the Dirac formalism reads $\sqsubseteq
  \varrho(\bfr,t), {\cal H}\sqsupseteq=0$.

The algebra needed to derive equation of motion for the velocity field
${\bf v}$ is slightly more complex than these leading to the
continuity  equation.

Following Dirac procedure we obtain:
\begin{eqnarray}\label{dirvelo1}
\frac{\partial \bfv(\bfr,t)}{\partial t}=\sqsubseteq
  \bfv(\bfr,t), {\cal H}\sqsupseteq=\{\bfv(\bfr,t), {\cal
  H}\}-\sum_{ab}\int d{\bf z} d{\bf z}'\{\bfv(\bfr,t),\Theta_{a}({\bf
  z})\}M_{ab}({\bf z},{\bf z}')\{\Theta_{b}({\bf z}'),{\cal H}\}\;.
  \end{eqnarray}
Evaluation of the right hand side of (\ref{dirvelo1}), with use of
expressions (\ref{Mab}) gives:
:
\begin{eqnarray}\label{dirvelo2}
  \frac{\partial \bfv(\bfr,t)}{\partial t}=
  \bfv(\bfr,t)\times(\nabla\times\bfv(\bfr,t))-\nabla_r\left[\int d{\bf z}
  G(\bfr-{\bf z})\nabla_z\cdot\left\{\bfv({\bf z})\times
   \left(\nabla\times\bfv(\bfr,t)\right) \right\}\right]\;.
\end{eqnarray}
Thus we have obtained nonlinear, nonlocal equation for the velocity
field known from previous work \cite{Kim,Barbara}.

The above exercise in the  Dirac brackets calculation provides a novel
formulation of the Euler incompressible fluid. The viscous fluid
equations can now easily be derived by replacing the Dirac brackets by
the metriplectic brackets discussed in \cite{Enz}. We can also use the
Dirac brackets as starting point in the perturbation theory in which
compressibility corrections are calculated. To do so one formally
associates small parameter $\kappa$ to the matrix elements $M_{ab}$
and expresses the Poisson brackets by the Dirac one. To the first
order in $\kappa$ the expression is identical to that in
(\ref{Dirbrac}) with reversed role of the Poisson and Dirac brackets.

In conclusion we have shown in the above that the Poisson brackets
formulation of the fluid dynamics can be used to derive the canonical
theory of the incompressible fluid following the Dirac prescription.
The application of this theory will be discussed in following
publication.\vskip0.4cm

\noindent {\bf Acknowledgments.-}\\
 We would like to thank Cyril Malyshev for contributing discussion
during the earlier stage of this work.                                                                     %


\appendix
\usecounter{A}
\begin{center} {\bf Appendix A}
\end{center}
\noindent\textbf{Matrix M:} \\ The matrix elements $M_{ab}$ satisfy
the following system of partial differential equations: 
\beq \label{M21M22,z}
  \nabla_\bfz \cdot \left[\nabla_\bfz M_{21}(\bfx,\bfz) +
        \frac{1}{\rr(\bfz)}\nabla_\bfz M_{22}(\bfx,\bfz)
        \times (\nabla\times\bfv(\bfz))\right ] = \delta(\bfx-\bfz) ,
\eeq 
\beq \label{M12M22,x}
  - \nabla_\bfx \cdot \left [\nabla_\bfx M_{12}(\bfx,\bfz) +
        \frac{1}{\rr(\bfx)}\nabla_\bfx M_{22}(\bfx,\bfz)
         \times (\nabla\times\bfv(\bfx)) \right ] = \delta(\bfx-\bfz) ,
\eeq
\beqn 
\label{M12,z} 
  \triangle_\bfz M_{12}(\bfx,\bfz) = - \delta(\bfx-\bfz) , \\
\label{M21,x}
  \triangle_\bfx M_{21}(\bfx,\bfz) = \delta(\bfx-\bfz) ,
\eeqn
\beq \label{M11M12,z}
  \nabla_\bfz \cdot \left [\nabla_\bfz M_{11}(\bfx,\bfz) +
        \frac{1}{\rr(\bfz)}\nabla_\bfz M_{12}(\bfx,\bfz)
        \times (\nabla\times\bfv(\bfz))  \right ] = 0 ,
\eeq 
\beq \label{M11M21,x}
  \nabla_\bfx \cdot \left [\nabla_\bfx M_{11}(\bfx,\bfz) +
        \frac{1}{\rr(\bfx)}\nabla_\bfx M_{21}(\bfx,\bfz) \times (\nabla\times\bfv(\bfx)) \right ] = 0 ,
\eeq
\beqn 
\label{M22,z}
    \triangle_\bfz M_{22}(\bfx,\bfz) = 0,\\
\label{M22,x}
    \triangle_\bfx M_{22}(\bfx,\bfz) = 0.
\eeqn

It is easy to check that these equations are satisfied by matrix elements given below:
\begin{eqnarray}
\label{Mab}
  M_{11}(\bfx,\bfz)&=&-\int
  d\bfx'G(\bfx-\bfx')\nabla_{\bfx'}\cdot\left[
  \frac{1}{\varrho(\bfx')}\nabla_{\bfx'}G(\bfx'-\bfz)
  \times(\nabla\times\bfv(\bfx'))\right]\equiv{\mathcal{M}}\{G\}\;,\nonumber\\
  M_{12}(\bfx,\bfz)&=&-M_{21}(\bfx,\bfz)=-G(\bfx-\bfz)\;,\nonumber\\
  M_{22}(\bfx,\bfz)&=&0\;,
\end{eqnarray}

\noindent{\bf Details of the Dirac brackets evaluation for the ideal fluid:}

Consider the Hamiltonian (\ref{Ham1}), the Dirac bracket 
$\sqb \rr(\bfx), H \sqp$ reads:
\begin{eqnarray}\label{rho-H-dirac}
   \sqb \rr(\bfx),H \sqp &=& \{ \rr(\bfx), H \} - \sum_{i,j}\int d\bfz_1 d\bfz_2 \:
      \left\{\rr(\bfx),\Theta_i (\bfz_1)\right\}
      M_{ij}(\bfz_1,\bfz_2)\left\{\Theta_j (\bfz_2), H\right\}
      \nonumber \\
   &=& \nabla_\bfx \cdot \vec{J}(\bfx) - \int d\bfz\: \triangle_\bfx M_{21}(\bfx,\bfz)
             \left[\nabla_\bfz \cdot \vec{J} (\bfz) \right]\nonumber\\
             &+&
             \triangle_\bfx M_{22}(\bfx,\bfz)
             \left[ \nabla_\bfz \cdot( \bfv(\bfz)\times(\nabla\times\bfv(\bfz)) )
    - \triangle_\bfz \left(\mu(\bfv,\varrho)  \right) \right]   =  0
    \;.
\end{eqnarray}

One sees immediately that the right hand side of (\ref{rho-H-dirac})
vanishes due to  (\ref{M21,x},\ref{M22,x},\ref{Mab}). Here
$\mathbf{J}=\varrho\bfv$ denotes the fluid particle current and
$\mu(\bfv,\varrho)=
 |\bfv|^2/2 +\partial f(\rr(\bfz))/\partial \varrho$
is the moving fluid chemical potential.

 The continuity equation is then

\begin{eqnarray}\label{theta1-constant}
   \frac{\pa}{\pa t} \rr(\bfx,t) = \sqb \rr(\bfx,t), H \sqp = 0\;,
\end{eqnarray}
as expected.

Evaluating Dirac bracket $\sqb v^i(\bfx), H \sqp$ we obtain:
\begin{eqnarray}
  \sqb v^i(\bfx), H \sqp &=& \left\{v^i(\bfx), H\right\} - \sum_{a,b}
           \int d\bfz_1 d\bfz_2 \:
           \left\{v^i(\bfx),\Theta_a(\bfz_1)\right\} M_{ab}(\bfz_1,\bfz_2)
                          \left\{\Theta_b(\bfz_2),
                          H\right\}\nonumber\\
         &=& A^{i}_0 - \sum_{a,b}  A^{i}_{ab} .
\end{eqnarray}

After straightforward but lengthy calculations we obtain:

\begin{eqnarray}
A^{i}_0 &=& \left\{v^i(\bfx), H\right\} = {[
\bfv(\bfx)\times(\nabla\times\bfv(\bfx)) ]}^i -
           \nabla^i \left[ \mu(\bfv,\varrho) \right]\;,\nonumber \\
A^{i}_{11}&=& - \nabla_{\bfx}^i \int d\bfz\: M_{11}(\bfx,\bfz)
                   \left[\nabla_\bfz\cdot \vec{J}(\bfz)\right]\;,\nonumber \\
A^{i}_{22} &=& 0 \;,\nonumber\\
   A^{i}_{12}
  &=&  \nabla_{\bfx}^i \int d\bfz\: G(\bfx-\bfz)
       \left\{ \nabla_{\bfz} \cdot [ \bfv(\bfz)\times(\nabla\times\bfv(\bfz)) ]
                      - \triangle_{\bfz}\left[ \mu(\bfv,\varrho)\right]
      \right\}\;,\nonumber \\
   A^{i}_{21} &=&  - \int  d\bfz\: \frac{1}{\rr(\bfx)}
        \left[\nabla_\bfx G(\bfx-\bfz) \times
        (\nabla\times\bfv(\bfx))\right]^i
       \left[\nabla_\bfz\cdot \vec{J}(\bfz)\right]\;,\nonumber\\
\end{eqnarray}
Using above, together with equations (\ref{M12,z}) and (\ref{M22,z})
we obtain:

\begin{eqnarray}
  \label{second2}
    \frac{\pa}{\pa t} \bfv(\bfx,t)
    &=& \bfv(\bfx)\times(\nabla\times\bfv(\bfx))\nonumber\\
    &+& \int d\bfz\: \left[ \nabla_\bfx M_{11}(\bfx,\bfz) +
      \frac{1}{\rr(\bfx)}\nabla_\bfx G(\bfx-\bfz)\times(\nabla\times\bfv(\bfx))
                  \right] \left[ \nabla_\bfz\cdot \vec{J}(\bfz) \right]
    \nonumber \\
   &-& \int d\bfz\: \left[ \nabla_\bfx G(\bfx-\bfz) \right]
     \left\{ \nabla_\bfz \cdot [ \bfv(\bfz)\times(\nabla\times\bfv(\bfz)) ]
      \right\}\; .
\end{eqnarray}

Acting on both sides of equation (\ref{second2}) with operator ${\bf div} =
\nabla_\bfx\cdot$,  and using equations (\ref{M12M22,x},\ref{M11M21,x}) one gets:
\begin{eqnarray}
  \label{theta2-constant}
     \frac{\pa}{\pa t} \left[ \nabla_\bfx \cdot \bfv(\bfx,t) \right]
     =  \sqb \nabla_\bfx \cdot \bfv(\bfx,t), H \sqp
     = \nabla_\bfx \cdot \sqb \bfv(\bfx,t), H \sqp
     = \nabla_\bfx \cdot \frac{\pa \bfv}{\pa t} (\bfx,t)  = 0 .
\end{eqnarray}

 Thus $\Theta_2(\bfx) = \nabla_\bfx \cdot \bfv(\bfx,t) \:$ is a constant of motions, as expected. 
Condition $\rr = \rr_0$ implies that $\nabla_\bfx \cdot \vec{J}(\bfx,t)\:$ is also a constant of
motions, and $\nabla_\bfx \cdot \vec{J}(\bfx,t)=0$.

Now, using equations (\ref{Mab},\ref{theta2-constant}) one easily sees that equation
(\ref{second2}) reduces to

\begin{eqnarray}
  \label{u-equation1}
  \frac{\pa \bfv(\bfx,t)}{\partial t} = \bfv(\bfx,t)\times (\nabla\times\bfv(\bfx,t)) -
  \nabla_\bfx \left[ \int d\bfz \: G(\bfx-\bfz) \:\: \nabla_\bfz \cdot
         [ \bfv(\bfz,t)\times(\nabla\times\bfv(\bfz,t)) ] \right]\;,
\end{eqnarray}
which is exactly the Euler equation for an ideal, incompressible fluid in its integral form.


\end{document}